\documentclass[11pt,a4paper]{article}
\usepackage{amssymb}
\usepackage{amsmath}
\usepackage{epsfig}

\newcommand{\ket}[1]{| #1 \rangle}
\newcommand{\bra}[1]{\langle #1 |}

\newcommand{\av}[1]{ \left\langle #1 \right\rangle}
\newcommand{\rme}[1]{\mathrm{e^{#1}}}
\newcommand{\intd}[1]{\,\mathrm{d} #1}
\newcommand{\fourier}{\mathcal{F}}
\newcommand{\ifourier}{\mathcal{F}^{-1}}

\begin{document}
\title{Momentum kicks due to quantum localization}
\author{A.J.Short\thanks{tony.short@qubit.org}\\ {\em Centre for Quantum Computation,} \\
{\em Clarendon Laboratory, University of Oxford,}\\ {\em Parks
Rd., OX1 3PU, UK}}
\date{}
\maketitle{}

\begin{abstract}
The momentum changes caused by position measurements are a central
feature of wave-particle duality. Here we investigate two cases -
localization by a single slit, and which-way detection in the
double-slit interference experiment - and examine in detail the
associated momentum changes. Particular attention is given to the
transfer of momentum between particle and detector, and the
recoil of the measuring device. We find that single-slit
diffraction relies on a form of `interaction-free' scattering,
and that an ideal which-way measurement can be made without any
back-reaction on the detector.
\end{abstract}

\section{Introduction}

It is well known that a measurement of position can change the
momentum distribution of a quantum object, yet the precise nature
of such changes, and the mechanism by which momentum is
transferred between particle and detector, remains the subject of
debate. One of the earliest concerns was that of Karl Popper, who
considered the momentum changes in an entangled pair when one of
the particles is localized \cite{popper}, and whose ideas have
recently provoked renewed discussion and experimentation
\cite{kim_and_shih, unnikrishnan, my_popper}. Other
investigations by Renniger \cite{renniger} and Dicke
\cite{dicke1, dicke2} have focused on the momentum changes when a
particle is \emph{not} detected in a certain region, and there
has also been debate over the momentum changes in a recently
proposed atom-maser interference experiment \cite{scully, storey}.

A more familiar example, which we consider in section
\ref{single_slit_sec}, is that of single-slit diffraction. Here
the slit itself provides a form of position measurement, and the
diffraction pattern in the far-field reveals the particle's
momentum distribution. A particle initially in a plane-wave state
will acquire a momentum spread on passing through the slit in
accordance with the Heisenberg uncertainty relation $\Delta x
\Delta p \geq \hbar/2$. In order to conserve total momentum, we
would expect a correlated change in the momentum distribution of
the slit, but it is unclear how the momentum is carried between
particle and slit. From a wave-perspective, the diffraction
pattern seems to be generated by that part of the wave which
\emph{does not} interact with the slit, passing straight through
the aperture. Momentum transfer, on the other hand, is a
particle-like feature, and seems most easily explained by the
action of forces at the slit edge. Investigating these two
aspects, we reveal the peculiar `interaction-free' nature of the
the diffraction process.

The effect of a position measurement on a particle's momentum
distribution is also crucial to the double-slit interference
experiment (section \ref{double_slit_sec}), in which the
interference fringes vanish if a successful measurement is made
of which slit the particle passed through \cite{feynman_slit}.
This famous example of wave-particle duality has formed the basis
for many discussions of quantum mechanics, most notably Einstein's
recoiling-slit experiment \cite{einstein_slit} and Feynman's
light microscope. In both cases, the loss of interference is
strongly linked to a transfer of momentum between the incident
particle and the measuring device, and it was originally argued
that momentum transfer and the Heisenberg uncertainty relation
play the key role in enforcing wave-particle duality.

However, more recent which-way detectors do not rely explicitly
on momentum transfer, such as the atom-maser system of Scully,
Englert and Walther \cite{scully}, or the spin-based system of
Schulman \cite{schulman}. Instead, which-way information is stored
in some internal degree of freedom which becomes entangled with
the spatial coordinate of the particle. A typical detection
process has the form
\begin{equation} \label{which_way_evolve_eqn}
\frac{1}{\sqrt{2}} ( \psi_A(x) + \psi_B(x)) \ket{0} \rightarrow
\frac{1}{\sqrt{2}} ( \psi_A(x) \ket{1} + \psi_B(x) \ket{0} )
\end{equation}
where $\psi_A(x)$ and $\psi_B(x)$ are wave packets just behind the
two slits and $\{\ket{0}, \ket{1}\}$ represent orthogonal states
of the detector. Because of this tagging process, any
interference between the two wavepackets is lost, yet either of
the wavepackets alone will remain completely unchanged by its
interaction with the detector (suggesting that the particle
experiences no momentum kick). Here it seems that entanglement,
rather than the uncertainty relation, is responsible for
enforcing wave-particle duality.

This approach has been criticised by Storey \emph{et al.}
\cite{storey}, who claim that momentum kicks and the uncertainty
relation are still crucial to the loss of interference. There can
be no doubt that the particle's momentum changes in the which-way
experiment, as the interference pattern obtained in the far field
\emph{is} a measurement of the particle's momentum distribution,
and it certainly changes when a detector is introduced (the
fringes disappear). We might therefore expect some momentum
transfer between the detector and particle, and a correlated
change in the detector's momentum distribution. In fact, we will
find in section \ref{double_slit_sec} that there is \emph{no}
change in the momentum distribution of the detector in such
cases. Despite this surprising result, we show that the total
momentum distribution remains unchanged in the interaction, as
required for momentum conservation.

As in the single-slit case, the change in particle momentum cannot
be attributed to classical forces, and the results strongly
suggest that the momentum change is an effect, and not a cause,
of the loss of interference. Similar situations are examined by
Wiseman \emph{et. al.} \cite{wiseman}, where it is shown that
they correspond to a `nonlocal' momentum transfer in the Wigner
function formalism.

\section{Single-slit diffraction} \label{single_slit_sec}

In this section, we consider the diffraction of a single
point-like particle by a rectangular slit, with a short-range
interaction between the particle and slit material close to the
slit edge. Both objects are treated quantum mechanically, and
their momentum changes are investigated.

\begin{figure}
\centerline{\includegraphics{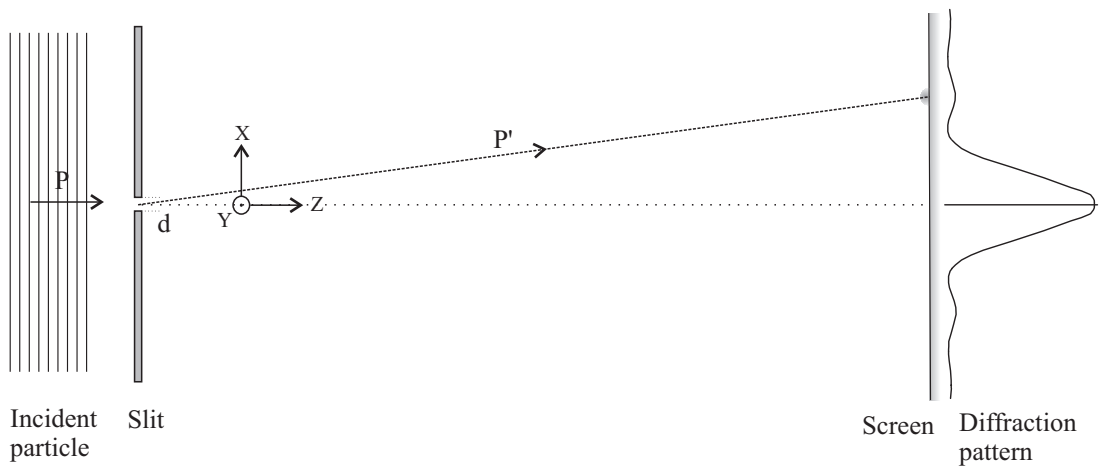}} \caption{The
single-slit diffraction setup, showing a possible scattering and
the corresponding particle momentum P'. In this case, we would
expect the slit to recoil in the $-x$ direction.} \label{figure_1}
\end{figure}

The diffraction setup is shown in figure \ref{figure_1}, in which
a single particle with a broad wavefunction and well-defined
momentum $\mathbf{P} \simeq (0,0,P)$ is normally-incident on a
narrow slit (of width $2d$). For simplicity, we restrict our
analysis to the $x$-direction in which the slit is narrow, and
model the particle's propagation in the $z$-direction by
comparing initial and final states on either side of the slit.
The initial particle state $\psi(x_p)$ is assumed to be constant
over the slit region, and given by
\begin{equation}
\psi(x_p) = \frac{1}{\sqrt{2L}}
\end{equation}
where $L (\gg d)$ is a measure of the width of the incident
wavefront.

The slit is assumed to be a rigid structure, with a gaussian
wavefunction $\phi(x_s)$ for the position of its centre,
\begin{equation}
\phi(x_s) = (2 \pi (\Delta x_s)^2)^{-1/4}  \exp \left(
\frac{-x_s^2}{4(\Delta x_s)^2} \right)
\end{equation}
Initially, we take $\Delta x_s \ll d$, giving the slit the sharp
localization we would expect of a macroscopic object. We also
assume that the slit remains relatively static during the
interaction, with a characteristic velocity spread $\Delta
\dot{x}_s$ which is much less than the particle velocity.

In reality, we would expect the slit to be in a thermal mixed
state, but this can always be decomposed into a probabilistic
mixture of pure states, and will not affect the results. A
typical value for $\Delta x_s$ can be calculated from the
equipartition theorem $\langle p_s^2/2M \rangle \sim
\frac{1}{2}kT$ and the gaussian uncertainty relation $\Delta x_s
\Delta p_s = \hbar/2$, giving
\begin{equation}
\Delta x_s \sim \left(\frac{\hbar^2}{4 M kT}\right)^{\frac{1}{2}}.
\end{equation}
For a slit of mass $M=10^{-3}$ kg at room temperature ($T=300$K),
this corresponds to a position uncertainty of $\Delta x_s \sim
10^{-23}$m and a velocity spread of $\Delta \dot{x}_s \sim 10^{-9}
\mathrm{m s}^{-1}$.

 We assume that the interaction between
the particle and slit is short-range, extending only a small
distance $\delta_f$($\ll d$) from the slit edge, and that it will
block any particles which collide directly with the slit
material. We model this by a potential distribution
\begin{equation}
V(x_p-x_s)  =  \left\{
\begin{array}{cl} V_0 & |x_p-x_s| \geq d+\delta_f  \\ V_s (x_p-x_s)
& d-\delta_f < |x_p-x_s| < d+\delta_f \\ 0 &|x_p-x_s| \leq
d-\delta_f \\ \end{array} \right.
\end{equation}
where $V_0$ is much greater than the kinetic energy of the
particle, and $V_s (x_p-x_s)$ is some smooth function taking the
potential between $V_0$ and $0$ close to the slit edge.

\begin{figure}
\centerline{\includegraphics{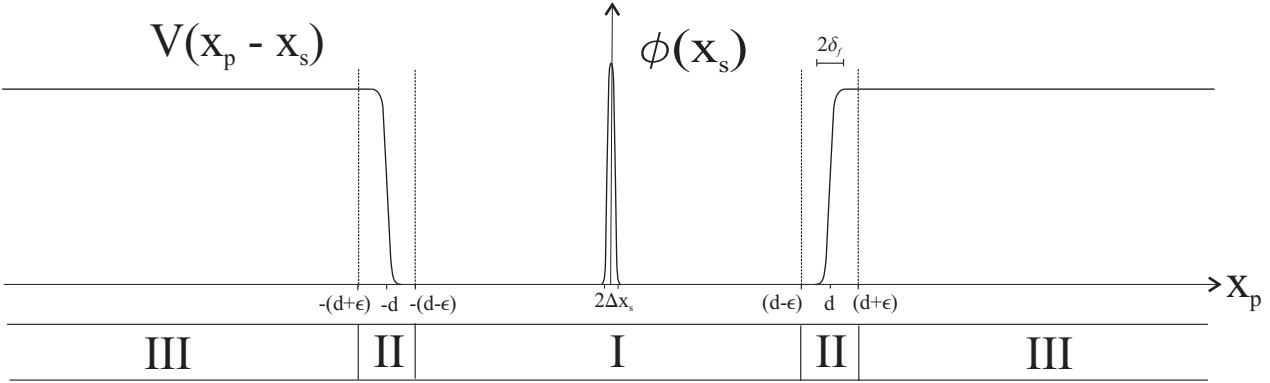}} \caption{The three
regions of $x_p$ into which the particle wavefunction is divided,
with the potential distribution $V(x_p-x_s)$ and the slit
wavefunction $\phi(x_s)$ shown on the same axis. The small
quantities $\epsilon, \Delta x_s$ and $\delta_f$ have been
exaggerated for clarity.} \label{slit_regions}
\end{figure}

To simplify the dynamics, we break the initial particle
wavefunction $\psi(x_p)$ into three parts $\{\psi_I(x_p),
\psi_{II}(x_p), \psi_{III}(x_p)\}$ which are unnormalized
projections onto the different spatial regions (fig.
\ref{slit_regions}). Region I ($|x_p| \leq  d - \epsilon$) lies
entirely within the slit aperture, where no forces act on the
particle. Region II ($d-\epsilon < |x_p| < d+\epsilon$) covers
the area close to the slit edge, where the forces generated by
$V_s (x_p-x_s)$ act on the particle, and region III ($|x_p| \geq
d+\epsilon$) covers the bulk of the slit material, where the
particle is blocked by the potential $V_0$.

To ensure that the component in region I passes freely through the
aperture, we set
\begin{equation}
\epsilon > (\Delta x_s +\delta_f + \delta_s),
\end{equation}
where $\delta_s (\ll d)$ is a measure of the transverse spreading
of the wavefunction during the interaction. We assume that the
slit is thin in the $z$-direction, and that the incident velocity
of the particle is much greater than its typical transverse
velocity ($P \gg h/d$) and that of the slit, such that the
particle travels rapidly through the slit with minimal spreading.
After this choice of $\epsilon$, any small residual interaction
with forces can be absorbed into the region II component.

We thus obtain an interaction of the form
\begin{eqnarray}
\psi_{I}(x_p) \phi(x_s) &\rightarrow& \psi_{I}(x_p) \phi(x_s) \label{slit_int1} \\
\psi_{II}(x_p) \phi(x_s) &\rightarrow& \Lambda_{II}(x_p,x_s) + \Gamma_{II}(x_p,x_s) \\
\psi_{III}(x_p) \phi(x_s) &\rightarrow& \Gamma_{III}(x_p,x_s)
\label{slit_int3}
\end{eqnarray}
where $\Lambda_{II}(x_p,x_s)$ is the entangled component at the
slit edge which passes through the aperture, and
$\Gamma_{II/III}(x_p,x_s)$ are components which are blocked by the
slit and do not contribute to the far-field interference pattern.
We leave off the slight spreading due to free evolution in the
region I component as this has no effect on the transverse
momentum distribution.

Under the influence of this interaction, the initial state
\begin{equation}
\Psi_i(x_p,x_s) = \psi(x_p)\phi(x_s) =
\left(\psi_I(x_p)+\psi_{II}(x_p)+\psi_{III}(x_s)\right)\phi(x_s)
\end{equation}
will evolve into an entangled state containing each of the terms
in equations (\ref{slit_int1})-(\ref{slit_int3}). We project out
only the component which has successfully passed through the slit
(which occurs with probability $\sim d/L$), to obtain the final
state
\begin{eqnarray}
\Psi_f(x_p,x_s) &=&  N \bigl(\psi_I(x_p)\phi(x_s) +
\Lambda_{II}(x_p,x_s)\bigr) \\
&=& \frac{N}{\sqrt{2L}} \, \chi_{d'}(x_p) \phi(x_s) + N
\Lambda_{II}(x_p,x_s),
\end{eqnarray}
where $N$ is a normalization constant ($\sim \sqrt{L/d}$) and
$\chi_{d'}(x_p)$ is a top-hat function which projects out the
wavefunction in region I, with value 1 if $|x_p| \leq
(d'=d-\epsilon)$ and 0 otherwise.

To obtain the far-field distribution, we Fourier-transform into
the momentum representation (with $\hbar=1$) to get
\begin{equation} \label{diff_patt}
\tilde{\Psi}_f(k_p,k_s)= \frac{N d'}{\sqrt{\pi L}}\left(
\frac{\sin(k_p d')}{k_p d'} \right) \tilde{\phi}(k_s) + N
\tilde{\Lambda}_{II}(k_p,k_s)
\end{equation}
Note that the entangled component $N
\tilde{\Lambda}_{II}(k_p,k_s)$, in which forces have acted
between the particle and slit, carries only a tiny fraction
($\sim \epsilon/d$) of the total probability for the state, and
the diffraction pattern will be largely determined by the first
term. Up to small corrections due to edge effects, we therefore
recover the familiar $\textrm{sinc}^2$ diffraction pattern in the
far-field, as given by the momentum probability distribution
\begin{equation}
\rm{Prob}(k_p) \simeq \left|N \tilde{\psi}_I(k_p)\right|^2 \simeq
\frac{d}{\pi} \left( \frac{\sin(k_p d)}{k_p d} \right)^2.
\end{equation}

However, as the slit wavefunction is unchanged in the first term,
only the small edge term $N \tilde{\Lambda}_{II}(k_p,k_s)$ can
contribute to any momentum change of the slit. This is what we
might expect classically, as forces are only present at the slit
edge, yet the particle's momentum distribution is largely
independent of these edge forces. Instead, the form of the
diffraction pattern is given by the region I contribution, in
which there has been no interaction between particle and slit.

Given that the changes in particle and slit momentum arise from
different terms, it is difficult to see how total momentum could
be conserved in the interaction. This is not necessarily
problematic, as we have projected out only part of the final
state (in which the particle passes through the slit), and the
total momentum distribution need only be conserved for the state
as a whole. Nevertheless, it would be a surprising result.

However, with the setup given above it would be impossible to
measure the momentum change of the slit. The typical momentum
transferred in the scattering ($\sim h/d$) will be far less than
the natural uncertainty in the slit momentum ($\hbar/2\Delta x_s$)
and hence, even with perfect recoil, the initial and final states
of the slit will be almost identical, with overlap very close to
1. It is this feature which allows for coherent reflection of
quantum particles from macroscopic objects such as
mirrors\cite{schulman}.

For the slit recoil to have a significant effect, the momentum
uncertainty of the slit must be similar in magnitude to (or less
than) the momentum changes due to diffraction, as deduced by Bohr
in his response to Einstein's recoiling-slit
experiment\cite{einstein_bohr}. To study the momentum changes of
the slit in more detail, we therefore consider a delocalized slit,
with $\Delta x_s \sim d$.

\begin{figure}
\centerline{\includegraphics{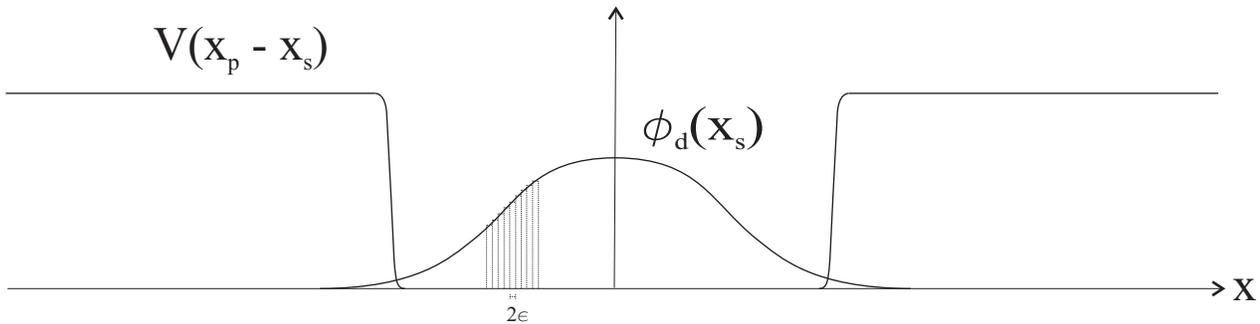}} \caption{The
delocalized slit wavefunction, which can be expressed as a
superposition of many localized states with width $2
\varepsilon$.} \label{delocalised_slit}
\end{figure}

With the position of the slit so uncertain, it is impossible to
define regions I-III for the state as a whole, but if we consider
the new state of the slit $\phi_d(x_s)$ as a superposition of
sharp wavefunctions at different positions  (fig.
\ref{delocalised_slit}), then we can evolve each term separately
as before. Dividing $\phi_d(x_s)$ into narrow strips of width $2
\varepsilon \ll d$, and then summing over $x'_s=  2 n
\varepsilon$ for all integer $n$ we have
\begin{equation}
\phi_d(x_s) \simeq \sum_{x'_s}
\chi_{\varepsilon}(x_s-x'_s)\phi_d(x'_s)
\end{equation}
and hence
\begin{equation}
\Psi_i(x_p, x_s) \simeq \sum_{x'_s} (\psi(x_p)
\chi_{\varepsilon}(x_s-x'_s)  ) \phi_d(x'_s).
\end{equation}
Evolving each bracketed term (in which the slit has the narrow
wavefunction $\chi_{\varepsilon}(x_s-x'_s)$) as before,
\begin{eqnarray}
\Psi_f(x_p,x_s) & = & \sum_{x'_s} N
 \left(\psi_{x'_s[I]}( x_p )\chi_{\varepsilon}(x_s-x'_s) + \Lambda_{x'_s[II]}(x_p,x_s)\right) \phi_d(x'_s)\\
& = & \sum_{x'_s} N \left( \frac{1}{\sqrt{2L}} \chi_{d'}(x_p -
x'_s) \chi_{\varepsilon}(x_s-x'_s) + \Lambda_{x'_s[II]}(x_p,x_s)
\right) \phi_d(x'_s).
\end{eqnarray}
We neglect the small contributions $\Lambda_{x'_s[II]}(x_p,x_s)$
from interactions close to the slit edge, and fourier-transform
to the momentum representation for the particle
\begin{eqnarray}
\tilde{\Psi}_f(k_p,x_s) &\simeq& \sum_{x'_s} \frac{N
d'}{\sqrt{\pi L}}\left( \frac{\sin(k_p d')}{k_p d'}\,
\mathrm{e}^{-i k_p x'_s} \right)
 \chi_{\varepsilon}(x_s-x'_s) \phi_d(x'_s)\\
& = &  \Biggl(\frac{N d'}{\sqrt{\pi L}} \left(\frac{\sin(k_p
d')}{k_p d'} \right) \Biggr)\Biggl(\sum_{x'_s}
 \left(\phi_d(x'_s)\mathrm{e}^{-i k_p x'_s} \right)\chi_{\varepsilon}(x_s-x'_s)
 \Biggr) \label{slit-eqn}
\end{eqnarray}
where we have used the fourier transform relation
$\mathcal{F}[\psi(x-a)] = \tilde{\psi}(k) \exp(-i k a)$. Finally,
we reconstruct the  slit wavefunction to give
\begin{equation} \label{recoileqn}
\tilde{\Psi}_f(k_p,x_s) \simeq \frac{N d'}{\sqrt{\pi L}}
\left(\frac{\sin(k_p d')}{k_p d'} \right)
\left(\phi_d(x_s)\mathrm{e}^{-i k_p x_s} \right)
\end{equation}
and hence
\begin{equation} \label{krecoileqn}
\tilde{\Psi}_f(k_p,k_s) \simeq N \tilde{\psi}_I(k_p)
\tilde{\phi}(k_s+k_p).
\end{equation}

The final state of the slit has the same gaussian form as the
initial state, but with an average momentum of $-k_p$, equal and
opposite to the particle momentum. We therefore obtain a simple
\emph{recoil} of the slit, with the momentum transfer
distribution given by $\mathrm{Prob}(k_p)$ for any slit
wavefunction. What is interesting is that the recoil derived here
has nothing to do with the forces at the slit edge. The
wavefunction given in equation (\ref{recoileqn}) was derived
entirely from region I components, in which the particle
propagates freely through the centre of the slit without any
interactions. In analogy with other recent works this appears to
be a form of `interaction free' scattering \cite{int_free}.

The momentum changes generated by the diffraction are a quantum
phenomenon, not a direct result of forces but of entanglement in
position and the intrinsic uncertainty in the slit position. Each
possible slit position $x_s$ generates a shifted diffraction
pattern, and for a given observed momentum $k_p$ this shift
corresponds to a phase change of $\exp(-i k_p x_s)$. As the
different slit positions add coherently we obtain a slit
wavefunction with an overall phase factor of $\exp(-i k_p x_s)$
and hence a momentum kick of $-k_p$. The role of momentum as a
translational symmetry property, rather than something which is
carried between particles by forces, is clearly emphasised.

This approach will be applicable whenever the particle moves
rapidly through the slit, and the slit wavefunction can be broken
into a large number of narrow strips which do not spread
significantly during the diffraction process. It can even be
applied to our original thermal slit, for which the initial
velocity spread $\Delta \dot{x}_s \sim 10^{-9} \mathrm{m s}^{-1}$
will probably be much less than that of the particle. The
wavefunction could therefore be divided into small strips with
 increased velocities which still wouldn't spread significantly
during the interaction, and the slight momentum kick due to
recoil could be explained by the above mechanism.

The apparent violation of energy conservation due to the
increased transverse momenta of particle and slit can be resolved
by including the $z$-component of the wavefunction in the
analysis. In addition to the transverse momentum, both
longitudinal momentum and energy are conserved in the extended
model (Appendix A), even though we have projected out only part of
the final state.

\section{Double-slit interference and which-way detection}
\label{double_slit_sec}

\begin{figure}
\centerline{\includegraphics{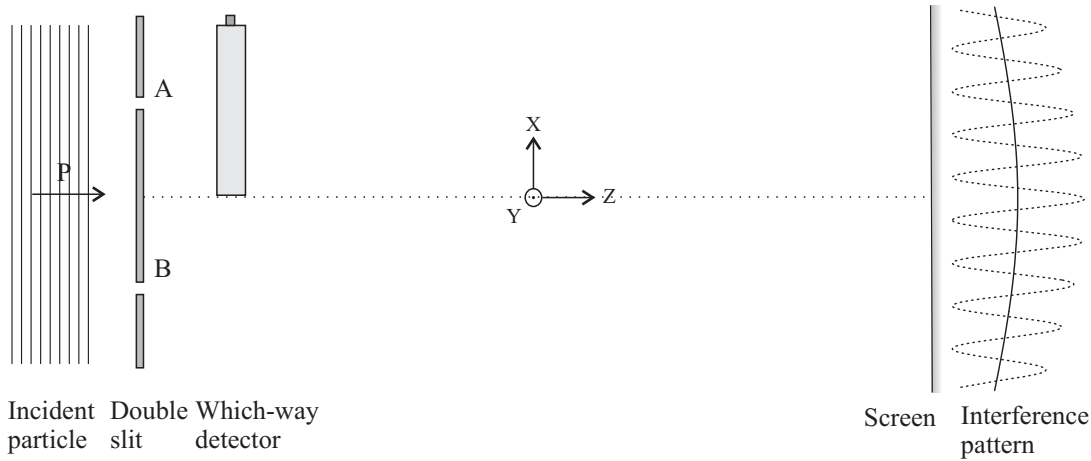}} \caption{Double slit
interference setup, showing the change in the diffraction pattern
(from dashed to solid plot) when the which-way detector is placed
between slit A and the screen.} \label{double_slit_fig}
\end{figure}

In this section we consider the effect of which-way detection on
the momentum distributions of the particle and detector in a
double slit interference experiment.
 The setup is shown in figure \ref{double_slit_fig},
in which a particle with well-defined momentum
$\mathbf{P}=(0,0,P)$ is incident normally on a screen with two
narrow slits. Just behind the upper slit (slit A) there is a
which-way-detector, the internal state of which will change from
$\ket{0}$ to $\ket{1}$ as the particle passes through it (as in
\cite{schulman}). Much further from the slits, in the far-field,
there is a sensitive screen which will register the position of
the particle, providing an effective measurement of its
transverse momentum. As the results in the previous section can
easily be extended to the double-slit itself, we focus our
attention on the action of the which-way detector.

As before, we restrict our analysis to the $x$-direction in which
the slits lie, and model the particle's propagation in the
$z$-direction by comparing initial and final states on either
side of the which-way detector. The particle state just before
the detector is
\begin{equation}
\psi(x_p) = \frac{1}{\sqrt{2}}(\psi_A(x_p) + \psi_B(x_p))
\end{equation}
where $\psi_A(x_p)$ and $\psi_B(x_p)$ are wavepackets behind
slits A and B respectively. We assume that $P$ is much greater
than the characteristic momentum of $\psi_A(x_p)$ and
$\psi_B(x_p)$, such that transverse spreading $\delta_s$ is
minimal while passing through the detection region.

\begin{figure}
\centerline{\includegraphics{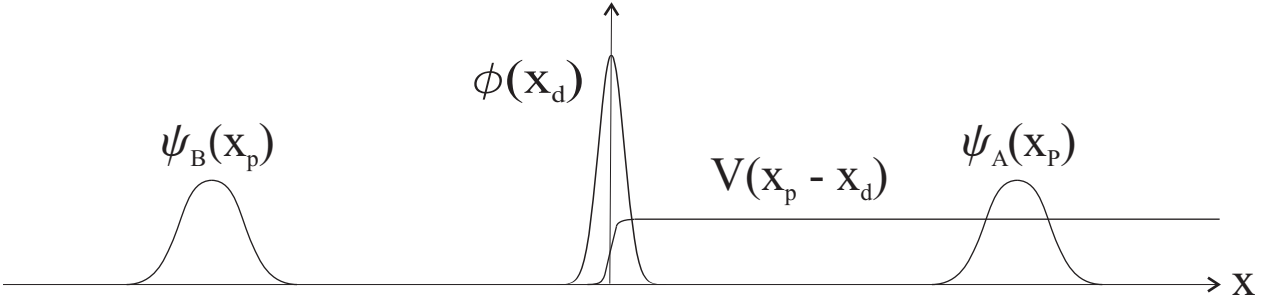}} \caption{The particle
wavefunction $\psi(x_p) = \frac{1}{\sqrt{2}}(\psi_A(x_p) +
\psi_B(x_p))$, detector wavefunction $\phi(x_d)$ and potential
distribution $V(x_p-x_d)$.} \label{double_slit_x_pic}
\end{figure}

The detector is a rigid well-localized object with a narrow
wavefunction $\phi(x_d)$ for the position of its left-hand edge
(which lies in the centre of the two slits). We choose an
interaction Hamiltonian of the form
\begin{equation} \label{interact_hamil_eqn}
H_I = -i V(x_p-x_d) (\ket{0}\bra{1} - \ket{1}\bra{0}) \quad
\mathrm{where}\quad V(x) = \left\{
\begin{array}{cl}
V_1 & x \geq \delta_f \\
V_d(x) & -\delta_f < x < \delta_f \\
 0 & x \leq -\delta_f
\end{array}
\right. ,
\end{equation}
with a constant potential $V_1$ within the detector and the edge
effects extending over a small region $\delta_f$ as before (fig.
\ref{double_slit_x_pic}). We set $V_1$ much lower than the kinetic
energy of the particle and assume that the potential is
smoothly-varying in the $z$-direction, such that almost all
particles pass through the detector without being reflected. By
appropriately choosing the thickness $w$ of the detection region
in the $z$-direction, we can ensure that the internal state of the
detector rotates completely from $\ket{0}$ to $\ket{1}$ as the
wavepacket $\psi_A$ passes through it (in time $\tau$). This
gives a unitary interaction of the form
\begin{equation}
U_I = \exp \left( \frac{-i H_I \tau}{\hbar} \right) = \left\{
\begin{array}{cl}
\left(\ket{1} \bra{0} - \ket{0}\bra{1} \right) & x \geq \delta_f \\
\left( \begin{array}{c} \sin \left(\frac{ \pi V_d(x)}{2 V_1}
\right) \left(\ket{1} \bra{0} - \ket{0}\bra{1}\right) \\ +\cos
\left(\frac{\pi V_d(x)}{2 V_1}\right) \left( \ket{0}\bra{0} +
\ket{1}\bra{1} \right) \end{array}
\right) & -\delta_f < x < \delta_f \\
  \left( \ket{0}\bra{0} +
\ket{1}\bra{1} \right) & x \leq -\delta_f
\end{array}\right.,
\end{equation}
where $x=x_p-x_d$, and we have assumed that there is no
possibility of particle reflection.

We assume that the wavepackets for the particle and detector
($\psi_A(x)$,$\phi(x)$,and $\psi_B(x)$) are all initially
separated by a distance greater than $\delta_f+\delta_s$, such
that they remain non-overlapping and free of edge effects
throughout the interaction, with support on different regions of
the $x$-axis. This allows the detector to provide a perfect
which-way measurement for the two slits, as given in equation
(\ref{which_way_evolve_eqn}). More precisely, the initial and
final states are given by
\begin{eqnarray}
\Psi_i(x_p, x_d) &=& \frac{1}{\sqrt{2}}(\psi_A(x_p) +
\psi_B(x_p))\phi(x_d) \ket{0}  \\ \Psi_f(x_p, x_d) &=&
\frac{1}{\sqrt{2}}(\psi_A(x_p)\ket{1} + \psi_B(x_p)\ket{0})
\phi(x_d),
\end{eqnarray}
where we have once again neglected the slight changes due to free
evolution during the detection process.

Because $\psi_A(x_p) \phi(x_d)$ always generates a positive
relative displacement $(x_p-x_d)>\delta_f$, and $\psi_B(y)
\phi(y_d)$ a negative displacement $(x_p-x_d)<-\delta_f$, the
interaction is not sensitive to the exact position of the
detector or the forces at the detector edge and hence the spatial
wavefunctions for the particle and detector remain unentangled.

In the regions of space $\{x_p, x_d\}$ in which the state is
non-zero, the interaction potential is always constant in the
$x$-direction, with
\begin{equation}
\frac{\partial V(x_p-x_d)}{
\partial x_p} = \frac{\partial V(x_p-x_d)}{
\partial x_d} = 0.
\end{equation}
Hence the particle and detector experience no transverse forces
during the detection process. Yet nevertheless the particle
momentum distribution $\mathrm{Prob}(k_p)$ is changed by the
interaction, from a distribution with interference fringes to one
without
\begin{equation}
\mathrm{Prob}_i(k_p) =
\left|\frac{1}{\sqrt{2}}\left(\tilde{\psi}_A(k_p) +
\tilde{\psi}_B(k_p)\right)\right|^2 \quad \rightarrow \quad
\mathrm{Prob}_f(k_p)
=\frac{1}{2}\left|\tilde{\psi}_A(k_p)\right|^2 +
\frac{1}{2}\left|\tilde{\psi}_B(k_p)\right|^2.
\end{equation}
The momentum distribution for the detector $\mathrm{Prob}(k_d)$,
however, remains completely unchanged, with
\begin{equation}
\mathrm{Prob}_i(k_d) = \mathrm{Prob}_f(k_d) =
\left|\tilde{\phi}(k_d)\right|^2
\end{equation}
Although simple, this result is actually quite surprising; not
only does the momentum of the particle change despite the absence
of forces, but there is an apparent violation of total momentum
conservation. Unlike the diffraction case there is no recoil of
the detector, the momentum distribution of the particle simply
changes on its own, \emph{without} any other change to balance it.

With the particle and detector unentangled, the total momentum
distribution $\mathrm{Prob}(k_T)$ is given by the convolution of
the two individual momentum distributions $\mathrm{Prob}(k_T)=
\mathrm{Prob}(k_p)\star \mathrm{Prob}(k_d)$. The translational
symmetry of the Hamiltonian (which depends only on the relative
coordinate ($x_p-x_d$)) should then ensure that
$\mathrm{Prob}(k_T)$ is conserved during the interaction.
Applying this to the above case, we require that
\begin{equation}
\left|\frac{1}{\sqrt{2}}\left(\tilde{\psi}_A(k_p) +
\tilde{\psi}_B(k_p)\right)\right|^2 \star
\left|\tilde{\phi}(k_d)\right|^2 =
\left(\frac{1}{2}\left|\tilde{\psi}_A(k_p)\right|^2 +
\frac{1}{2}\left|\tilde{\psi}_B(k_p)\right|^2\right) \star
\left|\tilde{\phi}(k_d)\right|^2.
\end{equation}
In fact, this equality will hold whenever $\phi(x)$ can fit into
the gap between $\psi_A(x)$ and $\psi_B(x)$(Appendix B), so total
momentum will be conserved in the detection process described
above. Note that this condition on the wavepackets is actually
necessary for the detector to be able to perfectly distinguish
the two particle states.

Nevertheless, this yields an interesting result: That the momentum
distribution of a single particle can change in isolation
\emph{without} changing the total momentum distribution. This is
a consequence of the convolution used to calculate the total
momentum distribution, which is not a one-to-one mapping. In this
case, the limited spatial support of the detector wavepacket
means that some information about the particle state is discarded
completely in the convolution. As the changes in particle
momentum only affect this discarded region, they cause no change
in the total momentum distribution.

We can also use this result to illustrate another feature of the
momentum changes: That, despite the obvious differences between
$\mathrm{Prob}_i(k_p)$ and $\mathrm{Prob}_f(k_p)$, the expectation
value of any finite positive integer power of the particle
momentum ($\langle k_p^N \rangle$ for $N\in\{0,1,2 \ldots\}$)
will be identical in the initial and final states. Using total
momentum conservation,
\begin{eqnarray}
\av{(k_p+k_d)^N}_i = \av{(k_p+k_d)^N}_f
\end{eqnarray}
where $\av{\,}_i$ and $\av{\,}_f$ represent expectation values in
the initial and final states respectively. Since the particle and
detector are unentangled (and uncorrelated), we can simplify the
power expansion to get
\begin{equation}
\sum_{n=0}^N \left( \frac{N!}{n!(N-n)!} \right) \av{k_p^n}_i
\av{k_d^{N-n}}_i  = \sum_{m=0}^N \left( \frac{N!}{m!(N-m)!}
\right) \av{k_p^m}_f \av{k_d^{N-m}}_f.
\end{equation}
Given that $\av{k_d^n}_i=\av{k_d^n}_f \; \forall \; n$, this
implies the inductive result that
\begin{equation}
\av{k_p^N}_i =\av{k_p^N}_f \;\; \rm{if} \;\;\; \av{k_p^m}_i
=\av{k_p^m}_f \;\; \forall \; m\in\{0,1,2,\ldots,N-1\}.
\end{equation}
As this equation is true for $N=0$ ($\av{k_p^0}_i
=\av{k_p^0}_f=1)$, it must therefore be true by induction for all
positive integer $N$. Thus $\rm{Prob}_i(k_P)$ and
$\rm{Prob}_f(k_P)$ are distributions with precisely the same
moments $(\av{k_p^N}_i =\av{k_p^N}_f)$. Those changes which do
occur can be best seen by considering the modular momentum $(k_p
\, \textrm{mod}\,\kappa)$, where $\kappa$ gives the fringe spacing
in the momentum distribution. This approach is followed in more
detail by Aharanov \emph{et al.} \cite{modular_momentum}.

\section{Discussion}

In both the single and double slit experiments, the momentum
changes associated with the position measurement cannot be viewed
as a direct result of classical forces, which vanish in the
relevant regions. Instead, they depend on quantum entanglement,
and the translational symmetries of the wavefunction.

In the single-slit case, we see the effect of a position
measurement inside a broad incident wavefront. Here the slit
provides a simple dual-valued position measurement, which tells us
whether or not the particle lies inside the aperture. The
uncertainty in the detector (slit) position yields a
superposition of final states, in which each detector position
state is entangled with a spatially-shifted particle state. In
momentum-space, these spatial-shifts of the particle correspond
to momentum-shifts of the detector, generating a recoil of the
detector which is just as as we would classically expect - even
though the scattering is `interaction-free'.

In the double-slit experiment, we consider the effect of a
position measurement which will distinguish spatially separated
parts of a wavefunction. In this case, an ideal detector can
perform the measurement without becoming spatially entangled with
the particle, or experiencing any recoil. Yet the momentum
distribution of the particle \emph{is} changed by the
interaction, even though it is not subject to any transverse
forces.

The fact that we can change the momentum distribution of the
particle without changing the momentum distribution of the
detector, and yet still conserve total momentum, is surprising.
However, it emerges as a natural consequence of the convolution
used to calculate the total momentum, in which some information
about the momentum distributions of the individual particles is
discarded.

I am grateful to Lucien Hardy, Lev Vaidman and W.Toner for useful
discussions, and to EPSRC for financial support.

\appendix

\section*{Appendix A}

The $z$-component of the wavefunction is important in determining
\emph{when} the particle passes through the aperture. Regions in
which the slit is closer to the incident particle ($z_s<0$) will
be diffracted sooner and thus experience a greater period of
transverse spreading.

To investigate this effect, we consider the wavefunction just
after the particle has passed through the slit, where it has the
approximate form
\begin{equation} \label{app_A_eqn}
\tilde{\Psi}_f(k_p, k_s, z_p, z_s) \simeq N \tilde{\psi}_I (k_p)
\tilde{\phi}(k_s+k_p, z_s) \rme{i P z_p / \hbar} ,
\end{equation}
where $\tilde{\phi}(k_s+k_p, z_s)$ now includes the slit
wavepacket in the z-direction, and the $x$-component is given in
terms of $\{k_p, k_s\}$ by equation (\ref{krecoileqn}). The slight
transverse spreading of the wavefunction during the interaction
period will introduce an additional phase factor, given by
\begin{equation}
\exp \left( \frac{- i E_x t}{\hbar} \right) =  \exp \left(-i
\left( \frac{\hbar k_{si}^2}{2 M} \right) t_i - i \left(
\frac{\hbar k_p^2}{2 m} + \frac{\hbar k_s^2}{2 M} \right) t_f
\right),
\end{equation}
where $k_{si}=k_s+k_p$ is the initial slit momentum before
diffraction, $t_i$ is the time at which the particle passed
through the slit and $t_f$ is the time since. The total time
$\tau= t_i+t_f$ is constant, but the proportion of the evolution
which occurs after diffraction will depend on the distance
between particle and slit. Taking the particle's speed in the
$z$-direction as $P/m$ and the slit as effectively static, we
approximate the evolution time after diffraction by
\begin{equation}
t_f \simeq \frac{m}{P} (z_p-z_s),
\end{equation}
which gives a phase factor of
\begin{equation}
\exp \left(-i \left( \frac{\hbar k_{si}^2}{2 M} \right) \tau  - i
\left( \frac{\hbar k_p^2}{2 P} + \frac{m \hbar (k_s^2 -
k_{si}^2)}{2 M P}\right) (z_p-z_s) \right).
\end{equation}
Including this phase factor in the final state wavefunction will
modify the $z$-momentum of the particle to
\begin{equation}
P' \simeq \left(1-\frac{\hbar k_p^2}{2 P^2} - \frac{m \hbar (k_s^2
- k_{si}^2)}{2 M P^2} \right) P
\end{equation}
with the slit receiving a momentum kick of equal magnitude in the
opposite direction. Note that the momentum changes in the
$z$-direction are much smaller than those in the $x$-direction,
due to our assumptions that $\hbar k_p \ll P$ and $m \ll M$. As
$(P'-P) \ll P$, the change in energy is approximately given by
\begin{equation}
\frac{{P'}^2}{2m}-\frac{P^2}{2m} \simeq \frac{P(P'-P)}{m} = -
\frac{\hbar k_p^2}{2 m} - \frac{ \hbar (k_s^2 - k_{si}^2)}{2 M},
\end{equation}
which exactly cancels the energy changes due to the increased
transverse momentum of the particle and slit. Thus, at the level
of these approximations, both energy and momentum are conserved
in the diffraction process.

\section*{Appendix B}

Proof that
\begin{equation} \label{app_eqn}
\left|\frac{1}{\sqrt{2}}\left(\tilde{\psi}_A(k) +
\tilde{\psi}_B(k)\right)\right|^2 \star
\left|\tilde{\phi}(k)\right|^2 =
(\frac{1}{2}\left|\tilde{\psi}_A(k)\right|^2 +
\frac{1}{2}\left|\tilde{\psi}_B(k)\right|^2) \star
\left|\tilde{\phi}(k)\right|^2.
\end{equation}
where $\star$ represents convolution and $\phi(x)$ is narrower
than the gap between $\psi_A(x)$ and $\psi_B(x)$:

Note that spatial translations on $\phi(x)$, and on $\phi_A(x)$
and $\phi_B(x)$ together, of the form
\begin{eqnarray} \phi(x) \rightarrow \phi(x-a) &\quad& \tilde{\phi}(k) \rightarrow
\mathrm{e}^{-i k a} \tilde{\phi}(k) \\
\psi_A(x) \rightarrow \psi_A(x-b) &\quad& \tilde{\psi}_A(k)
\rightarrow \mathrm{e}^{-i k b} \tilde{\psi}_A(k) \\
\psi_B(x) \rightarrow \psi_B(x-b) &\quad& \tilde{\psi}_B(k)
\rightarrow \mathrm{e}^{-i k b} \tilde{\psi}_B(k) \\
\end{eqnarray}
have no effect on equation (\ref{app_eqn}), thus without loss of
generality we consider the centre of $\phi(x)$ (of width $2w$),
and the centre of the gap between $\psi_A(x)$ and $\psi_B(x)$ (of
width $2v$ with $(v>w)$) to coincide at the origin, placing
$\phi(x)$ precisely in the centre of the gap.

Equation (\ref{app_eqn}) is true if and only if
\begin{equation} \label{app_eqn_2}
(\tilde{\psi}^{*}_A(k)\tilde{\psi}_B(k) +
\tilde{\psi}^{*}_B(k)\tilde{\psi}_A(k)) \star
(\tilde{\phi}^{*}(k)\tilde{\phi}(k)) =0.
\end{equation}
Using the Fourier-transform pair
\begin{eqnarray}
\mathcal{F}[\psi(x)] &=&  \frac{1}{\sqrt{2\pi}}\int \psi(x)
\rme{-i k x}
\intd x  = \tilde{\psi}(k) \\
\mathcal{F}^{-1}[\tilde{\psi}(k)]  &=& \frac{1}{\sqrt{2\pi}}\int
\tilde{\psi}(k) \rme{+ i k x} \intd k = \psi(x),
\end{eqnarray}
and the relations
\begin{eqnarray}
\mathcal{F}^{-1}\left[\tilde{\psi}_1(k) \star
\tilde{\psi}_2(k)\right] &=& \sqrt{2 \pi} \,
\mathcal{F}^{-1}\!\left[\tilde{\psi}_1(k)\right]
\mathcal{F}^{-1}\!\left[\tilde{\psi}_2(k)\right] \\
\mathcal{F}^{-1}\!\left[\tilde{\psi}^*_1(k) \tilde{\psi}_2 (k)
\right] &=& \frac{1}{\sqrt{2\pi}} \left(\psi_1^*(-x) \star \psi_2
(x)\right)
\end{eqnarray}
 we rewrite equation (\ref{app_eqn_2}) as
\begin{eqnarray}
&& \fourier \left[ \ifourier \left[
\left(\tilde{\psi}^{*}_A(k)\tilde{\psi}_B(k) +
\tilde{\psi}^{*}_B(k)\tilde{\psi}_A(k)\right) \star
\left(\tilde{\phi}^{*}(k)\tilde{\phi}(k)\right) \right] \right] =0 \\
&\Leftrightarrow& \fourier \left[ \left(\ifourier
\left[\tilde{\psi}^{*}_A(k)\tilde{\psi}_B(k)\right] + \ifourier
\left[ \tilde{\psi}^{*}_B(k)\tilde{\psi}_A(k) \right] \right)
\left(\ifourier \left[ \tilde{\phi}^{*}(k)\tilde{\phi}(k))
\right] \right) \right]=0 \\
&\Leftrightarrow& \fourier \left[\left( \left( \psi^{*}_A(-x)
\star \psi_B(x)\right) +  \left( \psi^{*}_B(-x) \star \psi_A(x)
\right) \right) \left( \phi^{*}(-x) \star \phi(x) \right) \right]
=0.
\end{eqnarray}

Note that the states $\psi^{*}_A(-x)$ and $\psi_B(x)$ both lie on
the same side of the origin in the region $|x|\geq v$. Their
convolution must therefore lie in the region $|x| \geq 2v$.
Applying the same argument to the states $\psi^{*}_B(-x)$ and
$\psi_A(x)$ we conclude that the first term in the product will
be non-zero only in the region $|x| \geq 2v$. However, the second
term in the product $( \phi^{*}(-x) \star \phi(x))$ can only be
non-zero in the region $|x| \leq 2w$. As $w<v$ the two functions
have no common support and their product (and its
Fourier-transform) will always equal zero, thus proving the
validity of equation \ref{app_eqn}.

\end{document}